\documentclass[aps,pra,10pt,twocolumn]{revtex4}

\usepackage{amsmath,amssymb}
\usepackage[T1]{fontenc}
\usepackage{lmodern}
\usepackage{microtype}
\usepackage{graphicx}
\usepackage{placeins}
\usepackage{mhchem}
\usepackage{hyperref}
\newcommand{\beq}{\begin{equation}}
\newcommand{\eeq}{\end{equation}}

\begin{document}

\title{Quantum interpretation of classical LiF simulations}

\author{Federico Brivio$^{1}$ and Luca Salasnich$^{1,2}$}
\affiliation{$^{1}$Dipartimento di Fisica e Astronomia ``Galileo Galilei'', Universit\`a di Padova, via Marzolo 8, I-35131 Padova, Italy
\\
$^{2}$INFN Sezione di Padova, via Marzolo 8, I-35131 Padova, Italy}

\begin{abstract}
Recent studies have shown that classical LiF simulations can reproduce low-temperature experimental spectra when performed at a simulation temperature $T^*$ higher than the physical temperature $T$. We show that the characteristic scale of this rescaling follows naturally from quantum vibrational statistics. An effective temperature $T_{\rm eff}(T)$ is obtained by matching classical and quantum harmonic energies and is evaluated using an analytic acoustic-optical model and a first-principles LiF vibrational density of states $D(\varepsilon)$. The resulting temperature reproduces the low-temperature scale of the reported spectral temperature and approaches the physical temperature in the classical regime. Our results identify zero-point motion and Planck-Bose-Einstein statistics as the origin of the reported temperature scale.
\end{abstract}

\maketitle

\section{Introduction}

At low temperature, the vibrational energy of a quantum crystal differs substantially from its classical counterpart because each phonon mode retains
a non-vanishing zero-point contribution \cite{PathriaBeale,Kittel}.
One may nevertheless associate a classical effective temperature with the quantum system by requiring equality of the total vibrational energies.
This construction depends on the complete phonon spectrum and, in a crystal with two atoms per primitive cell, must include both acoustic and
optical branches \cite{BornHuang,Kittel}.
It is therefore natural to ask whether an effective temperature obtained in this way can account for the temperature rescaling observed in classical simulations of infrared spectra.

Classical simulations of LiF infrared spectra have been reported in Refs.~\cite{EPL2015,PhysicaA2018} and subsequently discussed in
Refs.~\cite{CercZPE,Carati2026}. 
In these calculations the reflectivity is obtained from Newtonian trajectories through the Green-Kubo theory.
The agreement with the experimental spectrum at a physical temperature $T$ is obtained by performing the classical simulation at a different temperature $T^*$.
Here and throughout, the adjective ``classical'' refers to Newtonian dynamics and to the use of classical phase-space averages.
The interatomic interaction used in the simulations is an effective potential fitted to lattice-dynamical data. It incorporates microscopic electronic effects rather than being derived from a bare classical interaction \cite{PhysicaA2018}. 
At the two lowest temperatures, the values of $T^*$ were selected by trial and error to reproduce the measured spectral features
\cite{PhysicaA2018}.
In contrast, at all higher temperatures, no analogous adjustment was performed and $T^*$ was simply set equal to the physical temperature $T$. 

In this paper we test whether the characteristic scale of $T^*$ can be explained by the temperature $T_{\rm eff}$ of a classical harmonic crystal having the same internal energy as its quantum counterpart. We first introduce an analytic acoustic-optical model which makes the mode counting and the origin of the low-temperature energy scale transparent. We then remove its main spectral approximations by evaluating $T_{\rm eff}$ from a first-principles vibrational density of states (DOS). The comparison identifies the quantum vibrational energy scale underlying the reported temperature rescaling. Finally, for completeness, we use the same DOS to calculate the harmonic Helmholtz free energy, entropy, and constant-volume heat capacity of LiF comparing quantum vs classical predictions of statistical mechanics.

\section{Energy matching for harmonic oscillators}

The internal energy  of a quantum harmonic oscillation with frequency $\omega$ at temperature $T$ is given by 
\cite{PathriaBeale}:
\begin{equation}
U_Q(T,\omega) =
\frac{\hbar\omega}{2}\coth\!\left(\frac{\hbar\omega}{2k_BT}\right) =
\frac{\hbar\omega}{2} + \frac{\hbar\omega}{e^{\hbar\omega/(k_BT)}-1}.
\label{singlequantum}
\end{equation}
This expression contains both the zero-point contribution $\hbar\omega/2$ and the Planck-Bose-Einstein thermal contribution.
The corresponding classical oscillator has a mean energy
\begin{equation}
U_C(T_{\rm eff})=k_BT_{\rm eff}.
\end{equation}
Equating the two energies gives
\begin{equation}
T_{\rm eff}(T,\omega) =
\frac{\hbar\omega}{2k_B} \coth\!\left(\frac{\hbar\omega}{2k_BT}\right).
\label{singleTeff}
\end{equation}
The same relation follows by equating the quantum and classical mean-square position fluctuations.  It tends to $T$ in the classical high-temperature
limit and to $\hbar\omega/(2k_B)$ as $T\to0$.
Related frequency-dependent effective temperatures have been discussed in~\cite{Mine1,Mine2}.
For this single oscillator, $\omega$ is a fixed parameter.
In a crystal, it is replaced by a set of collective normal-mode frequencies determined by the dynamical matrix of the entire interacting lattice.

\subsection{Many-mode system}

Consider $M$ independent harmonic modes with frequency $\omega_j$.
Their classical and quantum internal energies are:
\begin{equation}
U_C(T_{\rm eff})=E_{\min}+Mk_BT_{\rm eff}
\label{classicalmany}
\end{equation}
and
\begin{equation}
U_Q(T)
=
E_{\min} +
\sum_{j=1}^{M}\frac{\hbar\omega_j}{2}\coth\!\left(\frac{\hbar\omega_j}{2k_BT}\right),
\label{quantummany}
\end{equation}
respectively.
Here $E_{\min}$ is the common equilibrium value of the potential energy and fixes an arbitrary energy origin.
It cancels when the two expressions are equated, yielding the global effective temperature
\begin{equation}
T_{\rm eff}(T)
=
\frac{1}{M}\sum_{j=1}^{M}
\frac{\hbar\omega_j}{2k_B}
\coth\!\left(\frac{\hbar\omega_j}{2k_BT}\right).
\label{globalTeff}
\end{equation}
Thus, $T_{\rm eff}$ is the average of the single-oscillator effective temperatures over all modes. It guarantees equality of the total energies, not equality of the energy of each individual mode.

\section{Analytic acoustic-optical model}

For an explicit treatment of the crystal modes, Eq.~\eqref{globalTeff} can be rewritten by replacing $j$ with the wave-vector and branch indices $({\bf q},s)$:
\begin{equation}
T_{\rm eff}(T) =
\frac{1}{M}\sum_{{\bf q},s}\frac{\hbar\omega_s({\bf q})}{2k_B}\coth\!\left[\frac{\hbar\omega_s({\bf q})}{2k_BT}\right].
\label{crystalTeff}
\end{equation}
Here ${\bf q}$ is the wave vector and $s$ is the phonon-branch label. 
The frequencies $\omega_s({\bf q})$ are collective properties of the crystal: they depend on the interactions, masses, lattice structure, branch index, and wave vector.
The primitive cell of LiF contains two atoms, giving rise to six phonon branches: three acoustic and three optical. Each group contains two transverse branches and one longitudinal branch.
All six branches contribute to Eq.~\eqref{crystalTeff}.

The exact dispersions depend on both the magnitude and direction of ${\bf q}$.
To obtain an explicit result, we replace the first Brillouin zone with a Debye sphere and approximate the three anisotropic acoustic branches by three degenerate isotropic branches whose frequencies depend only on $|{\bf q}|$.

For the acoustic modes, we use the sinusoidal dispersion
\begin{equation}
\omega_a({\bf q})=
\frac{2c_s q_D}{\pi}
\sin\!\left(\frac{\pi |{\bf q}|}{2q_D}\right),
\qquad |{\bf q}|\leq q_D,
\label{SinusoidalDispersion}
\end{equation}
while the three gapped optical branches are taken to be degenerate and dispersionless:
\begin{equation}
\omega_o({\bf q})=\omega_0.
\label{omega0simple}
\end{equation}

We assume a crystal containing $N$ primitive cells. For each acoustic branch,
the Debye replacement is
\begin{equation}
\sum_{\bf q} \longrightarrow
\frac{V}{(2\pi)^3}\int_{|{\bf q}|\leq q_D}d^3q =
\frac{V}{2\pi^2}\int_0^{q_D} dq \, q^2 .
\label{DebyeMeasure}
\end{equation}
The cutoff is fixed by requiring $N$ wave vectors per branch:
\begin{equation}
\frac{V}{(2\pi)^3}\frac{4\pi q_D^3}{3}=N.
\label{DebyeCutoff}
\end{equation}

The angular integration already includes the directions ${\bf q}$ and $-{\bf q}$; no additional factor of two must be introduced.

The normalization in Eq.~\eqref{SinusoidalDispersion} preserves the sound velocity at long wavelength: $\omega_a({\bf q})=c_s|{\bf q}|+O(q^3)$.
Compared with a linear Debye spectrum, it also represents the flattening of acoustic branches as they approach the boundary of the Brillouin zone.

We define
\begin{equation}
\Theta_a=\frac{\hbar c_s q_D}{k_B},
\qquad
\Theta_0=\frac{\hbar\omega_0}{k_B}.
\end{equation}
Introducing $x=|{\bf q}|/q_D$, the mode temperature associated with the
acoustic dispersion is
\begin{equation}
\theta_a(x)=\frac{\hbar\omega_a(xq_D)}{k_B}
=\frac{2\Theta_a}{\pi}\sin\!\left(\frac{\pi x}{2}\right).
\label{AcousticModeTemperature}
\end{equation}
The quantum energies of the three acoustic and three optical branches are, respectively,
\begin{equation}
U_a(T)=3Nk_B\,{\cal T}_a(T),
\label{AcousticEnergy}
\end{equation}
with 
\begin{equation} 
{\cal T}_a(T) =
\frac{3}{2}\int_0^1 dx\,x^2\theta_a(x) \coth\!\left(\frac{\theta_a(x)}{2T}\right),
\end{equation}
and
\begin{equation}
U_o(T)=3Nk_B\,{\cal T}_0(T),
\label{OpticalEnergy}
\end{equation}
with
\begin{equation} 
{\cal T}_0(T) =
\frac{\Theta_0}{2}
\coth\!\left(\frac{\Theta_0}{2T}\right).
\end{equation}

Since the crystal has $6N$ modes, its classical vibrational energy is $6Nk_BT_{\rm eff}$.  The condition $6Nk_BT_{\rm eff}=U_a+U_o$ therefore gives
\begin{equation}
\begin{aligned}
T_{\rm eff}(T)={}&
\frac{3}{4}\int_0^1 dx\,x^2\theta_a(x)
\coth\!\left(\frac{\theta_a(x)}{2T}\right) \\
&+\frac{\Theta_0}{4}
\coth\!\left(\frac{\Theta_0}{2T}\right).
\end{aligned}
\label{DebyeEinsteinTeff}
\end{equation}

Two limits provide useful consistency checks. At zero temperature ($T=0$) one obtains 
\beq
T_{\rm eff}(0) =
\frac{12(\pi-2)}{\pi^4}\Theta_a+\frac{1}{4}\Theta_0 ,
\label{ZeroTemperatureTeff} 
\eeq
whereas at high temperature one gets $T_{\rm eff}(T)\to T$.

For LiF, which has two atoms per primitive cell, the acoustic cutoff corresponding to the tabulated Debye temperature is $\Theta_a=581\ {\rm K}$ \cite{JonesWard2016}.
For the effective optical frequency in Eq.~\eqref{omega0simple}, we use the representative value of $\omega_0=8.69$ THz ($5.46\times10^{13}\ {\rm rad\,s^{-1}}$ as reported in Ref. ).
It gives $\Theta_0=\hbar\omega_0/k_B=417\ {\rm K}$. 
Equation~\eqref{ZeroTemperatureTeff} then implies  $T_{\rm eff}(0)=186$ K.

\begin{figure}[ht]
  \centering
\includegraphics[width=\columnwidth]{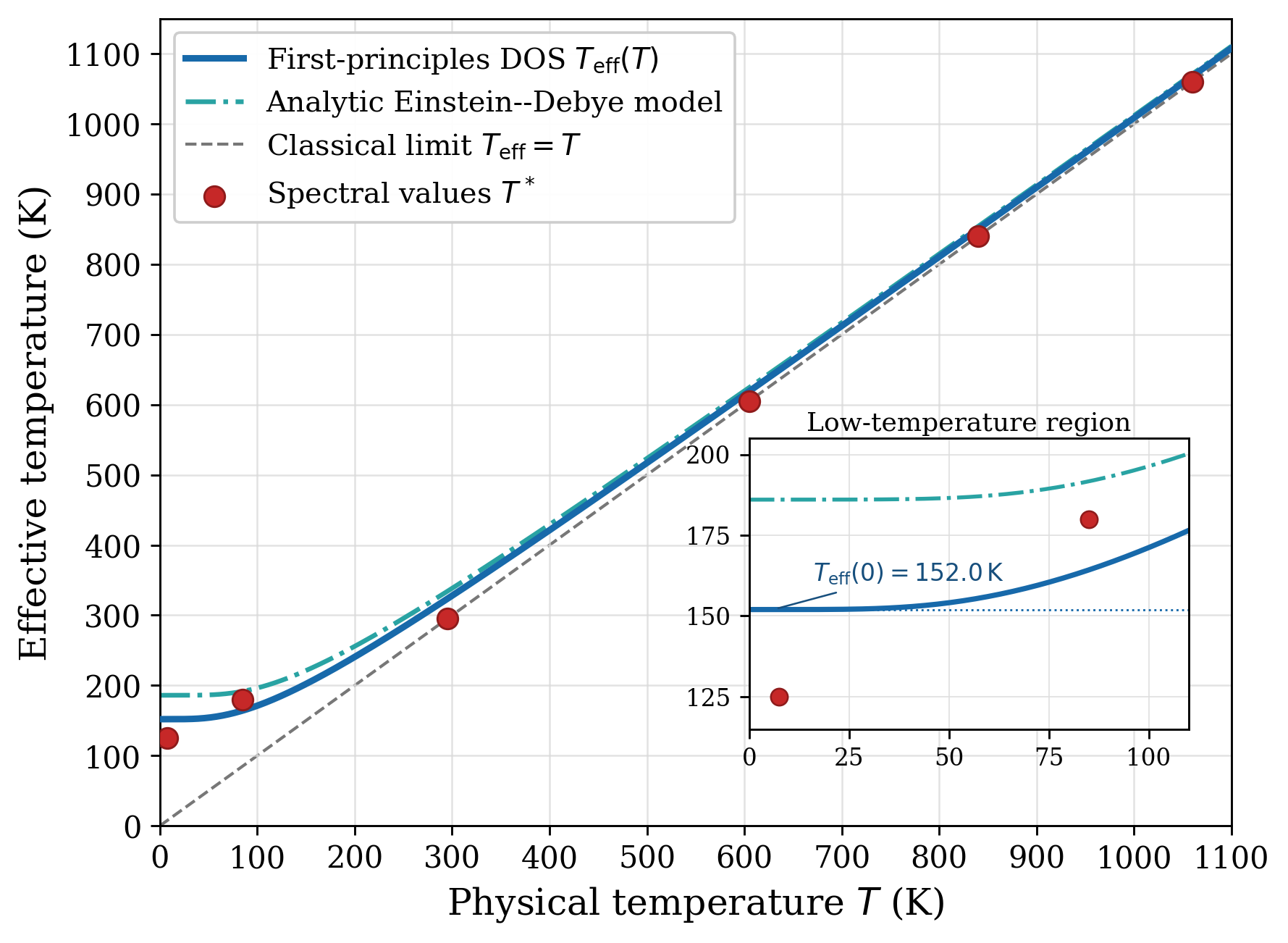}
  \caption{Energy-matching effective temperature obtained from the first-principles vibrational DOS (blue solid line), compared with the
  analytic acoustic-optical model (dash-dotted line) and the spectral   temperatures $T^*$ used in the classical simulations of LiF (red filled
  circles) \cite{EPL2015,PhysicaA2018,CercZPE,Carati2026}.  
  The gray dashed line denotes the classical limit $T_{\rm eff}=T$. 
  The inset enlarges the low-temperature region and shows the first-principles limiting value   $T_{\rm eff}(0)=152.0$ K.}
  \label{fig:teff}
\end{figure}

As shown in Fig.~\ref{fig:teff}, the simple Einstein-Debye model (dash-dotted line) predicts the correct classical limit and a finite quantum plateau. With the representative parameters above it gives $T_{\rm eff}=186$ K at $7.5$ K and $192$ K at $85$ K, compared with
$T^*=125$ K and $180$ K (filled circles). 
To test how sensitive this comparison is to the assumed phonon spectrum, we replace the model with the full vibrational density of states, as discussed in the next section.

\section{First-principles vibrational density of states}

The analytic acoustic-optical model provides a transparent description of the phonon spectrum. A more general treatment uses the vibrational density of states (DOS), defined as
\begin{equation}
D(\varepsilon)
=
\sum_{{\bf q},s}
\delta\!\left[\varepsilon-\hbar\omega_s({\bf q})\right]
\label{VibrationalDOS}
\end{equation}
where $\varepsilon_s({\bf q})=\hbar\omega_s({\bf q})$. The DOS is normalized according to
\begin{equation}
\int_0^\infty d\varepsilon \,  D(\varepsilon) =M.
\label{DOSnormalization}
\end{equation}
Here $M$ is the total number of vibrational modes. For a three-dimensional system containing $N_{\rm at}$ atoms, $M=3N_{\rm at}$. The DOS can be measured experimentally~\cite{Sturhahn1995} or computed from first principles~\cite{YinCohen1982}.

\subsection{Density functional perturbation theory}
In this work, we calculated the DOS using density-functional perturbation theory (DFPT)~\cite{Baroni2001} applied to the equilibrium primitive cell of LiF (space group $Fm\bar{3}m$, No.~225).
The computational details are given in Appendix~\ref{app:computational}.

Figure~\ref{fig:dos} shows the resulting DOS per primitive cell as a function of the vibrational energy $\varepsilon=h\nu=\hbar\omega$.

\begin{figure}[ht]
  \centering
\includegraphics[width=\columnwidth]{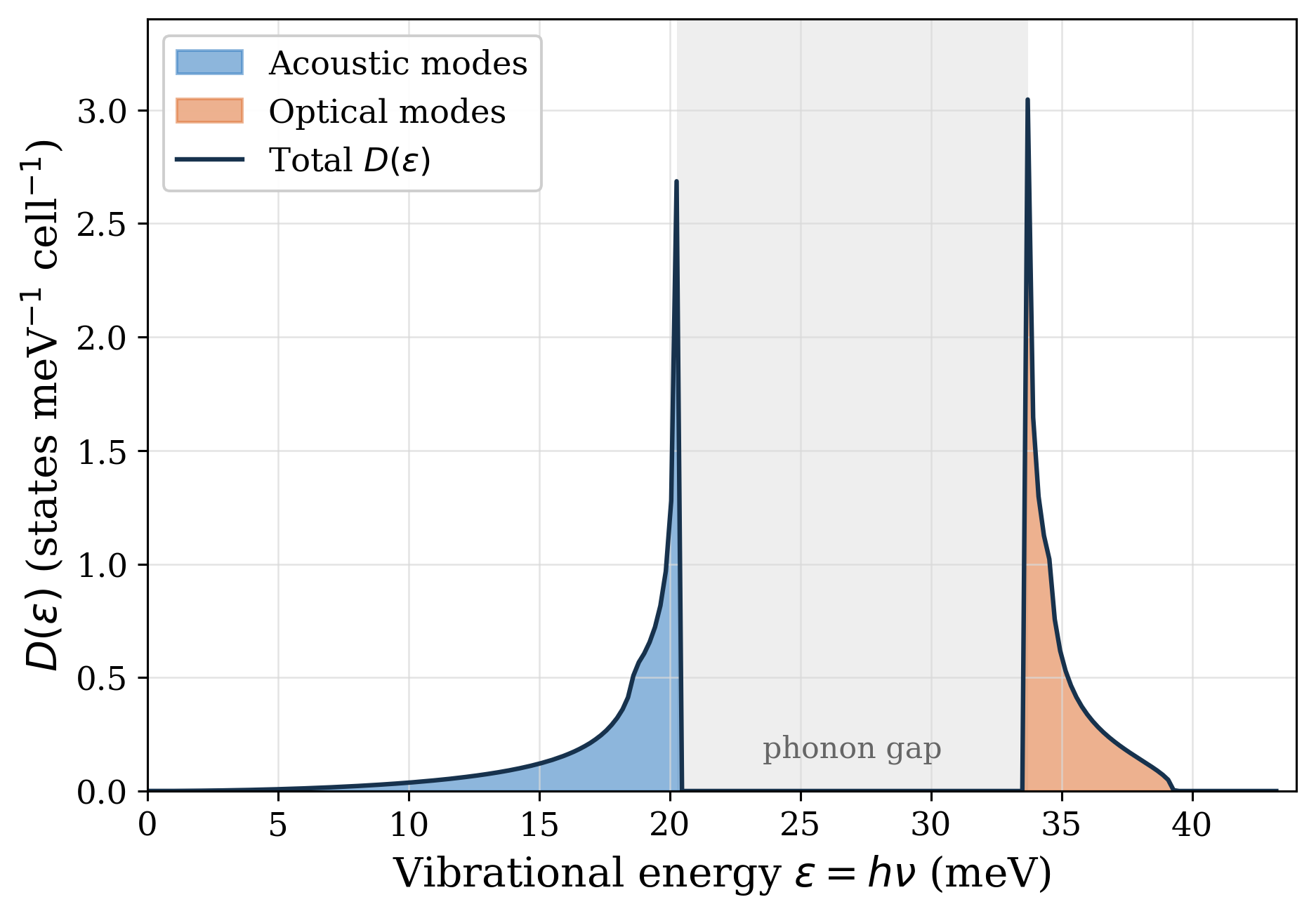}
  \caption{First-principles vibrational density of states of LiF per
  primitive cell. The two groups (acoustic and optical) are separated by a phonon gap.}
  \label{fig:dos}
\end{figure}

In terms of $D(\varepsilon)$, the quantum vibrational energy measured from the equilibrium minimum is

\begin{equation}
U_Q(T)-E_{\min} =
\int_0^\infty d\varepsilon\,D(\varepsilon)\frac{\varepsilon}{2} \coth\!\left(\frac{\varepsilon}{2k_BT}\right).
\label{QuantumEnergyDOS}
\end{equation}
The corresponding classical energy is
\begin{equation}
U_C(T_{\rm eff})-E_{\min}=Mk_BT_{\rm eff}.
\end{equation}
Equating the two energies gives
\begin{equation}
T_{\rm eff}(T)=
\frac{
\displaystyle
\int_0^\infty d\varepsilon\,D(\varepsilon)
\frac{\varepsilon}{2k_B}
\coth\!\left(\frac{\varepsilon}{2k_BT}\right)
}{
\displaystyle
\int_0^\infty d\varepsilon\,D(\varepsilon)
}.
\label{TeffDOS}
\end{equation}
If the density of states is given per primitive cell and normalized according to
\begin{equation}
\int_0^\infty d\varepsilon \, D(\varepsilon) =6,
\end{equation}
Eq.~\eqref{TeffDOS} reduces to
\begin{equation}
T_{\rm eff}(T) =
\frac{1}{6}
\int_0^\infty d\varepsilon\,D(\varepsilon)
\frac{\varepsilon}{2k_B}
\coth\!\left(\frac{\varepsilon}{2k_BT}\right).
\label{TeffDOSperCell}
\end{equation}
At zero temperature, Eq.~\eqref{TeffDOS} gives
\begin{equation}
T_{\rm eff}(0)
=
\frac{1}{2k_B}
\frac{
\displaystyle\int_0^\infty
d\varepsilon\,\varepsilon D(\varepsilon)
}{
\displaystyle\int_0^\infty
d\varepsilon\,D(\varepsilon)
},
\label{TeffZeroDOS}
\end{equation}
so that the low-temperature plateau is determined by one half of the mean phonon energy, expressed in temperature units.  At high temperature,
\begin{equation}
T_{\rm eff}(T)
=
T+
\frac{\langle\varepsilon^2\rangle_D}{12k_B^2T}
+O(T^{-3}),
\label{TeffHighTDOS}
\end{equation}
where
\begin{equation}
\langle\varepsilon^2\rangle_D
=
\frac{
\displaystyle\int_0^\infty
d\varepsilon\,\varepsilon^2D(\varepsilon)
}{
\displaystyle\int_0^\infty
d\varepsilon\,D(\varepsilon)
}.
\end{equation}
Thus, the classical limit $T_{\rm eff}(T)\to T$ is recovered independently of the detailed form of the density of states.

For the selected DOS, normalized to six modes per primitive cell, the acoustic and optical sectors contain $3.060$ and $2.940$ modes, respectively. Their mean frequencies are $4.31$ THz and $8.44$ THz, while the mean over all six branches is $6.33$ THz, corresponding to $\langle\varepsilon\rangle_D=26.2\ {\rm meV}$. Equation~\eqref{TeffZeroDOS} then predicts $T_{\rm eff}(0)=152.0\ {\rm K}$. This value is substantially below the $186$ K obtained with the illustrative parameters of the analytic model and is considerably closer to the lowest reported spectral value, $T^*=125$ K.

The full temperature dependence obtained from Eq.~\eqref{TeffDOS} is shown by the solid line in Fig.~\ref{fig:teff}. The dash-dotted line is the analytic acoustic-optical result of Eq.~\eqref{DebyeEinsteinTeff}, retained to illustrate the effect of replacing the simplified spectrum with the calculated DOS. The first-principles DOS lowers the predicted low-temperature saturation scale from $186$ K to approximately $152$ K, thereby markedly improving the comparison with the lowest-temperature spectral value. At $85$ K, the calculated value is $T_{\rm eff}=164.3$ K, about $15.7$ K below the reported $T^*=180$ K. At higher temperatures, $T_{\rm eff}$ progressively approaches the physical temperature, consistent with Eq.~\eqref{TeffHighTDOS}.

The first-principles DOS includes the anisotropy and nondegeneracy of the six branches and the optical dispersion within the adopted electronic-structure approximation. It therefore removes the principal spectral assumptions of the simplified model. Nevertheless, the DOS is obtained from a first-principles calculation of LiF, whereas the classical simulations use a fitted effective interatomic potential. For a strictly homogeneous comparison, an additional DOS should be obtained by harmonic linearization of that same potential. Comparing the two densities of states would quantify how faithfully the simulation potential reproduces the vibrational spectrum used in the present quantum-statistical construction.

\subsection{Thermodynamic functions}

Once the vibrational density of states $D(\varepsilon)$ is known, several thermodynamic quantities, including the Helmholtz free energy, entropy, and the constant-volume heat capacity, can be straightforwardly evaluated within the harmonic approximation.

The vibrational Helmholtz free energy is
\begin{equation}
\begin{aligned}
F_{\rm vib}(T)={}&
 \int_0^\infty d\varepsilon \,  D(\varepsilon)
\biggl[\frac{\varepsilon}{2} \\
&\quad+k_{\rm B}T
 \ln\!\left(1-e^{-\varepsilon/(k_{\rm B}T)}\right)
 \biggr] .
\end{aligned}
\end{equation}
The zero-point contribution is therefore included in $F_{\rm vib}$.

The vibrational entropy and the constant-volume heat capacity follow from $S_{\rm vib}=-(\partial F_{\rm vib}/\partial T)_V$ and $C_V=T(\partial S_{\rm vib}/\partial T)_V$. Introducing $x=\varepsilon/(k_BT)$, they are
\begin{equation}
 S_{\rm vib}(T)=k_{\rm B}\int_0^\infty d\varepsilon \,  D(\varepsilon)
 \left[
 \frac{x}{e^x-1}-\ln\left(1-e^{-x}\right)
 \right] 
\end{equation}
and
\begin{equation}
 C_V(T)=k_{\rm B}\int_0^\infty d\varepsilon \, D(\varepsilon)
 \frac{x^2e^x}{(e^x-1)^2} .
\end{equation}
Molar quantities are obtained by multiplying these expressions by
Avogadro's number. We notice for the last equation the high-temperature limit
$C_V\rightarrow 6R$ per mole of LiF.

\begin{figure}[ht]
  \centering
\includegraphics[width=0.90\columnwidth]{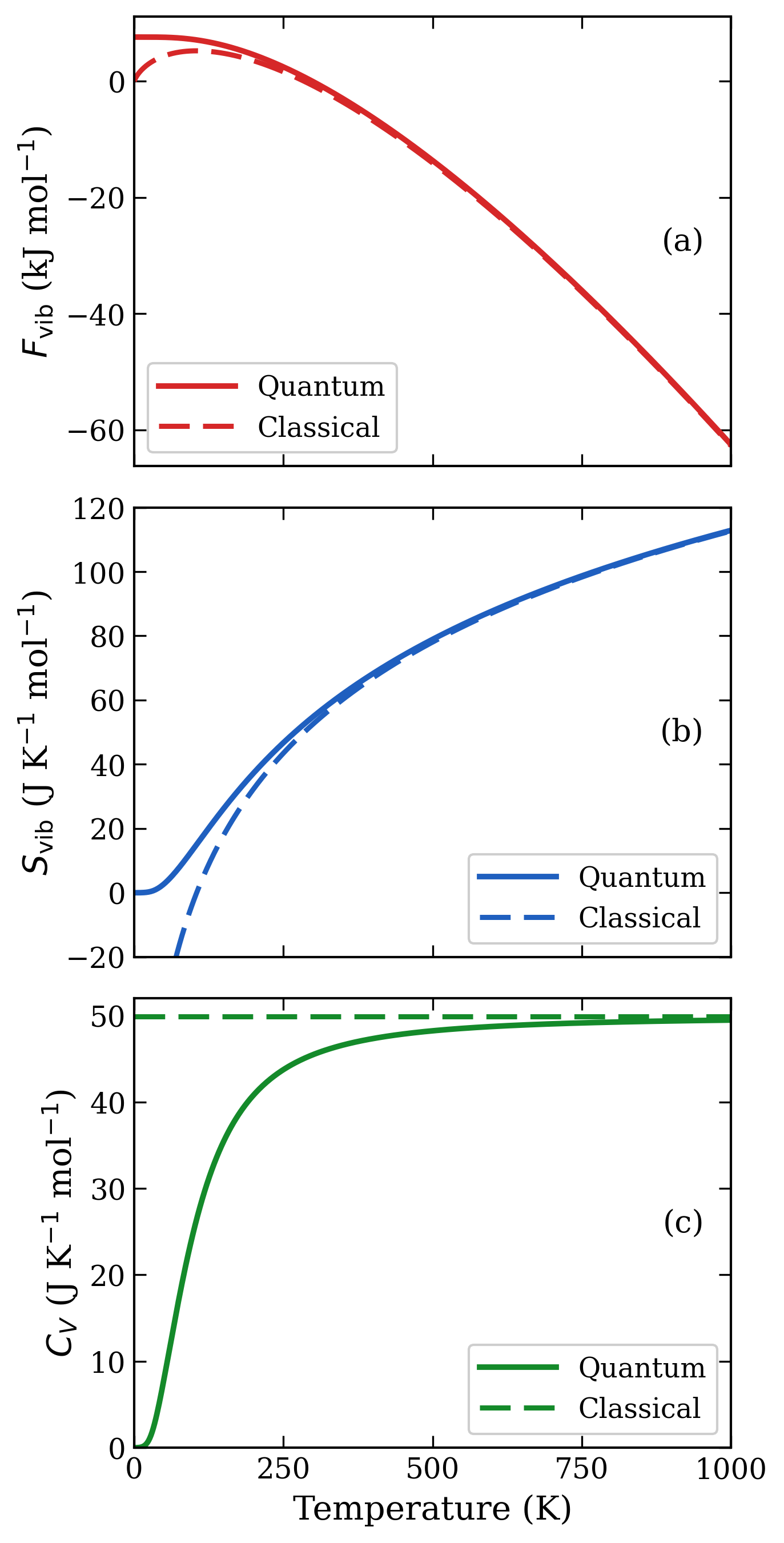}
\caption{Quantum (solid lines) and classical canonical (dashed lines) harmonic thermodynamic functions of LiF obtained from the first-principles vibrational DOS normalized to six modes per primitive cell: (a) Helmholtz free energy, (b) entropy, and (c) constant-volume heat capacity. The zero-point contribution is included in the quantum free energy. The classical phase-space normalization is chosen so that the classical expressions reproduce the high-temperature limits of the quantum results. Classically, $C_V=6R$ at all temperatures, whereas the entropy diverges to $-\infty$ as $T\to0$; consequently, the vertical axis in panel (b) has been truncated.}
  \label{fig:thermal}
\end{figure}

The quantities obtained for LiF are shown in Fig.~\ref{fig:thermal} as solid lines. At zero temperature, both the entropy and the heat capacity vanish, whereas the Helmholtz free energy reduces to the zero-point energy. As the temperature increases, the progressive thermal population of the vibrational modes produces a monotonic increase in entropy and a corresponding decrease in free energy. The heat capacity rises rapidly at low and intermediate temperatures and gradually approaches the classical Dulong-Petit limit, $C_V=6R$ per mole of LiF. These limits provide consistency checks on the normalization of the DOS and its use in the harmonic thermodynamic functions. For comparison, Fig.~\ref{fig:thermal} also shows, in all three panels, the corresponding results obtained from classical statistical mechanics (dashed lines). The quantum and classical predictions progressively approach each other as the temperature increases. However, their low-temperature behaviors are qualitatively different. In the classical description, the free energy does not contain a zero-point contribution, the entropy diverges to negative infinity as $T\to0$, and the heat capacity remains equal to $6R$ at all temperatures. Quantum statistics, instead, gives a finite $F_{\rm vib}(0)$ that is the zero-point vibrational energy, while both $S_{\rm vib}$ and $C_V$ vanish as $T\to0$, consistent with the third law of thermodynamics. The comparison therefore makes explicit the importance of nuclear quantum effects in the low-temperature regime.

\FloatBarrier

\section{Conclusions}

We have shown that 
equating the classical and quantum internal energies defines a unique global effective temperature $T_{\rm eff}$ for a harmonic crystal. The analytical model developed here contains three acoustic and three optical branches with the correct mode counting and approaches the physical temperature $T$ in the classical limit. The first-principles DOS removes the assumptions of acoustic isotropy, branch degeneracy, and a dispersionless optical frequency. It gives $T_{\rm eff}(0)=152.0$ K, between the analytical estimate of $186$ K and the lowest reported \cite{EPL2015,PhysicaA2018,CercZPE,Carati2026}
spectral value $T^*=125$ K. At $T=85$ K it gives $T_{\rm eff}=164.3$ K, compared with the assigned value $T^*=180$ K. The comparison suggests that the characteristic spectral temperature reported for LiF follows the quantum vibrational energy scale determined independently from the phonon DOS. 
As an additional consistency check, the same normalized DOS yields vibrational thermodynamic functions with the expected zero-temperature behavior and the correct high-temperature Dulong-Petit limit.

The quantitative comparison can be refined in two ways. 
First, the procedure by which the numerical values of $T^*$ are extracted from the simulated spectra \cite{EPL2015,PhysicaA2018,CercZPE,Carati2026} is not described in sufficient detail to establish their uniqueness or statistical uncertainty.
Second, the comparison relies on low-temperature reflectivity spectra measured in 1966 \cite{Jasperse1966}, for which a modern independent
assessment of possible experimental and systematic uncertainties is not available.
New low-temperature measurements, together with an explicit quantitative fitting protocol for $T^*$, would enable a more stringent test of this interpretation. The present calculation clearly indicates that the relevant low-temperature energy scale follows from zero-point motion and Planck-Bose-Einstein statistics, without requiring a separate classical zero-point energy reservoir or attributing it solely to the stochasticity-threshold scenario proposed in Refs.~\cite{EPL2015,PhysicaA2018,CercZPE,Carati2026}.

\section{Acknowledgments}

The authors thank Andrea Carati and Luigi Galgani for useful and inspiring discussions. 
This work is partially supported by the Project ``Frontiere Quantistiche'' (Departimenti di Eccellenza) of the Italian Ministry of University and Research. F.B. acknowledges ISCRA (\href{https://userdb.hpc.cineca.it/node/39985}{IsB32\_INT3R}) for awarding this project access to the LEONARDO supercomputer, owned by the EuroHPC Joint Undertaking, hosted by CINECA (Italy). L.S. 
acknowledges partial support by ``Iniziativa Specifica Quantum'' of INFN, by the European Union-Next Generation EU within the European Quantum Flagship Project ``PASQuanS 2'', and the National Center for HPC, Big Data and Quantum Computing (Project No.~CN00000013, CN1 Spoke 10: ``Quantum Computing'').

\appendix
\section{Computational details}
\label{app:computational}

The calculations were performed with VASP using the projector-augmented-wave (PAW) method \cite{KresseFurthmuller1996a,KresseFurthmuller1996b,KresseJoubert1999} and the PBEsol exchange-correlation functional \cite{Perdew2008}. The plane-wave cutoff was 700 eV, and a $7\times7\times7$ ${\bf k}$-point mesh was used for the primitive cell.
Upon geometric relaxation we found a cubic cell parameter of $a=4.0145055$ \AA, equal to a cell lattice for the primitive cell of:
\[
\frac{a}{2}
\begin{bmatrix}
0 & 1 & 1 \\
1 & 0 & 1 \\
1 & 1 & 0
\end{bmatrix}.
\]
Li and F were kept at their crystallographic sites in the $Fm\bar{3}m$ structure
, with Li at $(0,0,0)$ and F at $(0.5,0.5,0.5)$. Using the DFPT, we obtained an optical-mode frequency at $\Gamma$ of $9.33$ THz (wavenumber $311$ cm$^{-1}$), in good agreement with the wavenumber experimental value of $315$ cm$^{-1}$ \cite{Eldridge1972}. Repeating the calculation with a denser $14\times14\times14$ ${\bf k}$-point mesh changed this frequency by less than $0.02$ meV. 

The vibrational data were analyzed with Phonopy \cite{Togo2023}. The DOS was generated using an $81\times81\times81$ ${\bf q}$-point mesh and a frequency spacing of $0.05$ THz. Its raw numerical integral is $5.924$ modes per primitive cell; all DOS-based results reported in the main text use the global normalization factor $6/5.924$ required by Eq.~\eqref{DOSnormalization}.

\end{document}